# Selective Rotational Excitation of Molecular Isotopes and Nuclear Spin Isomers

**Sharly Fleischer, I.Sh. Averbukh and Yehiam Prior**

*Department of Chemical Physics, Weizmann Institute of Science, Rehovot 76100, Israel*

[yehiam.prior@weizmann.ac.il](mailto:yehiam.prior@weizmann.ac.il), phone +972-8-934-4008, fax +972-8-934-4126

Following excitation by a strong ultra-short laser pulse, molecules develop coordinated rotational motion, exhibiting transient alignment along the direction of the laser electric field, followed by periodic full and fractional revivals that depend on the molecular rotational constants. In mixtures, the different species undergo similar rotational dynamics, all starting together but evolving differently with each demonstrating its own periodic revival cycles. For a bimolecular mixture of linear molecules, at predetermined times, one species may attain a maximally aligned state while the other is anti-aligned (i.e. molecular axes are confined in a plane perpendicular to the laser electric field direction). By a properly timed second laser pulse, the rotational excitation of the undesired species may be almost completely removed leaving only the desired species to rotate and periodically realign, thus facilitating further selective manipulations by polarized light. In this paper, such double excitation schemes are demonstrated for mixtures of molecular isotopes (isotopologues) and for nuclear spin isomers.

---------------------------------------------------------------------------------------------------------------------

Alignment and orientation of molecules have always intrigued spectroscopists, and provided a wide range of topics to be studied. In the gas phase, molecular alignment following excitation by a strong laser pulse was observed in the seventies[1], and proposed as a tool for optical gating. In the early experiments, picosecond laser pulses were used for the excitation, and deviation of the refractive index from that of an isotropic gas was monitored as evidence for alignment[2,3]. More recently, these observations have been revisited both theoretically and experimentally (for a recent review, see ref.[4]). Spatial and temporal dynamics was studied[5,6,7], and multiple pulse sequences giving rise to the enhanced alignment were suggested,[8,9] and realized[10,11,12,13]. Further manipulations such as the optical molecular centrifuge and alignment-dependent strong field ionization of molecules were demonstrated[13,14,15]. Molecular phase modulators have been shown to compress ultrashort light pulses[16,17] and molecular alignment has been used for controlling high harmonic generation[18,19,20]. Other experiments were reported where transient grating techniques were used for detailed studies of molecular alignment and deformation. [21,22].

Analysis and separation of molecular species with similar chemical and physical properties continue to be challenging pursuits in physics and chemistry[23]. Since separation is based on differences in one or more physical or chemical properties, selective manipulation is becoming increasingly more difficult as the species become more and more alike. In the case of isotopologues (chemical species that differ only in the isotopic composition of their molecules), the difference lies in the nuclear mass of the different species (and therefore also in their moment of inertia and rotatinal constant), whereas for spin isomers, the mass is the same, and the difference is in the response to an externall magnetic field and in the symmetry properties of the entire molecular wavefunction describing the state.

Here we introduce a new approach to selective manipulation of multi-component molecular mixtures. By employing *non-resonant* laser fields, we induce drastic transient contrasts between the angular distributions of the various species. These differences should eventually lead to their discrimination and separation.

We start with a detailed description of the molecular evolution following the application of a laser pulse, and we focus our attention on the coherent rotational motion manifested in a transient angular distribution.

When molecules with a permanent dipole moment (such as HCl, KI etc) are subject to a DC electric field, they align parallel to the applied field direction. The torque exerted on a molecule is proportional to the component of the field perpendicular to the molecular axis. A related effect is achieved with molecules **without a permanent dipole moment** such as $I_2$, $CO_2$, $N_2$ etc. When a laser pulse acts on such a molecule, it induces an electric dipole through the anisotropic molecular polarizability tensor, which in turn interacts with the electric field. This second order interaction aligns the molecules along the external field as well.

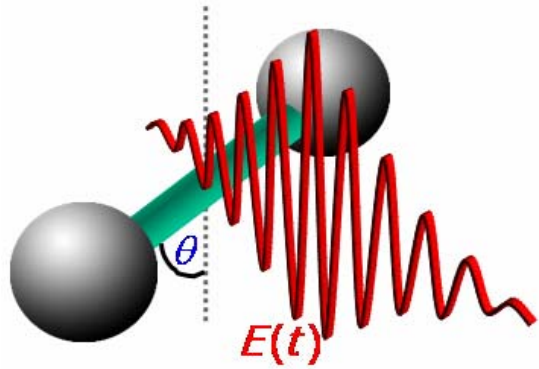

Figure 1: A laser pulse interacting with diatomic homonuclear molecule. The torque exerted on the molecule is proportional to $-\sin(2\theta)$, where $\theta$ is the angle between the laser pulse polarization and the molecular axis.

Let us consider an ensemble of polarizable molecules in gas phase, possessing isotropic initial angular distribution. A typical rotational time for a small molecule is of the order of several picoseconds, therefore with femtosecond pulse duration, the molecule can be regarded as frozen during the pulse. When a strong femtosecond laser pulse is applied, each molecule gains angular velocity depending on the angle $\theta$ between the molecular axis and the laser pulse polarization direction (see Fig. 1). The molecules start rotating towards the electric field direction immediately after the pulse, however a significant reorientation will be noticed only after the pulse is over, i.e. under field-free conditions[24]. During the pulse, every molecule experiences a torque which is proportional to the square of the projection of the field on the molecular axis. Assuming no initial rotational velocity prior to the pulse, a molecule at the angle $\theta$ will gain the angular velocity $\omega(\theta) \propto -P\sin(2\theta)$ (where $P$ can be considered as the "strength" of the pulse given by $P = \int_{-\infty}^{\infty} \varepsilon(t)dt$). According to this argument, a molecule which is oriented at $\theta = \pi/4$ to the field, will gain the highest angular velocity towards the field direction, whereas a molecule oriented perpendicular to the field ($\theta = \pi$), will gain no angular velocity. After the pulse, the molecules rotate according to their new angular velocity and short time after the pulse (several hundreds of femtoseconds) the distribution will have a peak in the direction of the field, a state that we refer to as the "*aligned state*" or "cigar". This state is manifested in the increase of the refractive index for light polarized in the direction of the exciting pulse[25].

Quantum-mechanically, a laser pulse excites a wave packet which is formed from many rotational states, and which may be written as:

$$|\psi(\theta,t)\rangle = \sum_{J,m} c_{Jm} |J,m\rangle \exp(-iE_J t/\hbar) \qquad (1)$$

For linear molecules, the energy spectrum is given by $E_J = hBc\, J(J+1)$ (where $B = \hbar/4\pi Ic$ is the rotational constant, $c$ is speed of light and $I$ is the molecular moment of inertia). As can be seen from Eq.(1) any such wave packet exactly reproduces itself at integer multiples of the revival time, $T_{rev} = 1/(2Bc)$ (quantum revivals phenomenon[26,27,28]). Shortly before the realignment revival, the molecules attain a state of "anti-alignment", namely a state where the molecular axes are confined in the plane perpendicular to the direction of alignment. In what follows, we refer to this "anti-aligned" state, as "disk" shaped, which is also accompanied by a reduction in the refractive index in the direction of the exciting pulse. Thus, every revived cigar-like angular distribution is preceded in time by a disk-like state (shown in Figure 2). In addition, it can be shown[8] that around the half-revival time ($t \approx T_{rev}/2$) the aligned and anti-aligned states appear in the reversed order. In between these special times, the molecules lose their macroscopic alignment and assume an almost isotropic angular distribution until the next revival event.

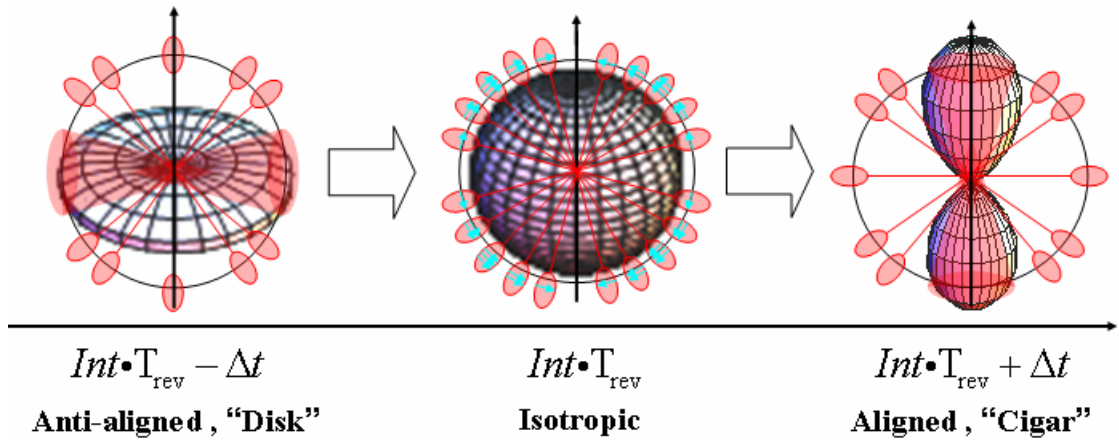

| $Int \cdot T_{rev} - \Delta t$ | $Int \cdot T_{rev}$ | $Int \cdot T_{rev} + \Delta t$ |
| Anti-aligned, "Disk" | Isotropic | Aligned, "Cigar" |

Figure 2: Angular distribution evolution around full revival times. The anti-aligned "Disk" state appears just before the revival time, and evolves through the isotropic state towards the aligned "Cigar" state. The classical 2D distribution (red) is overlapped with the 3D calculated angular distribution at the background. The small blue arrows correspond to the molecular rotational velocities induced by the pulse.

During its evolution, the rotational wave packet also exhibits other partially aligned and anti-aligned distributions at quarter and three quarters of the revival time. These are not as pronounced as the full and half revival times, due to internal destructive interferences between the odd and even rotational states.

## Experimental

We use a time-delayed degenerate, forward propagating, three dimensional, phase matched four-wave mixing arrangement[29], where the first two pulses set up a spatial grating of transiently aligned molecules, and the third, delayed pulse is scattered off this grating. In these experiments, all three input beams (and therefore the fourth output beam as well) were linearly, vertically polarized. The experiments were carried out at room temperature with 70 fs, $200 \mu J$ pulses from a regeneratively amplified Ti:Sapphire laser at 800 nm. The peak field intensity in the focal region was $3*10^{13}\ W/cm^2$. Under these conditions, the rotational energy supplied to a nitrogen molecule by the laser pulse is comparable with (and even exceeds) the thermal rotational energy. Based on procedures described earlier[8,30], we estimate the maximal degree of laser-induced alignment to be $<\cos^2\theta> \approx 0.5$ (compared to the isotropic 0.33)  The best antialignment is estimated to be of the order of $<\cos^2\theta> \approx 0.2$ . These estimations are consistent with the results of Refs.[11,15], in which a comparable degree of $N_2$ alignment was directly measured (using Coulomb explosion imaging) under experimental conditions (pulse energy, duration and focusing) generally similar to ours.

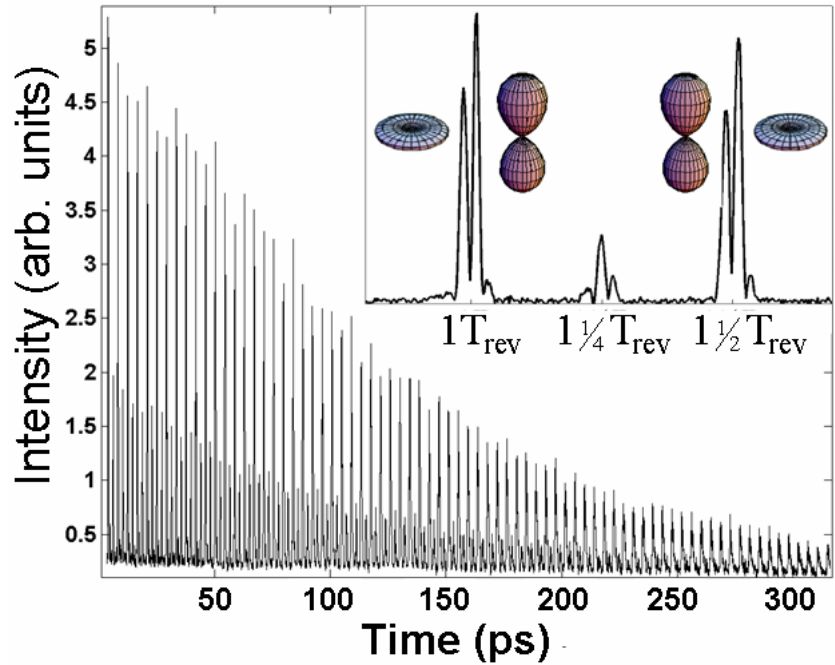

Figure 3: Long scan FWM signal from N$_2$ gas (200 torr, room temperature). 40 full revival cycles are shown (~320ps), where quarter, half and full revivals are clearly seen. The inset depicts one full revival cycle. Although the FWM signals around full and half revival seem similar, the corresponding angular distributions are reversed [8]

Figure 3 depicts the time delayed degenerate FWM signal obtained from a single isotope of nitrogen ($^{14}N_2$) following strong, ultrashort excitation. Over 40 revival cycles, 8.3 ps each, are observed, demonstrating full, half and quarter revivals. The overall decay of the revivals results from collisions within the cell, and the flight of molecules across the laser beam.

Next we consider control of molecular alignment by a pair of pulses. A single short laser pulse "kicks" the molecules, thus generating a rotational wave packet. However, in a double pulse excitation, the response of the molecular ensemble is very sensitive to the timing of the second pulse. If a second pulse is applied **at the time of exact revival**, its effect is similar to that of the first pulse, namely it kicks the molecules "in phase" with their rotational motion, and **adds angular momentum** to the already rotating molecules, resulting in a more pronounced alignment. This situation is illustrated by the measurements presented in figure 4a. When the second pulse is applied exactly at $3T_{rev}$, the amplitude of the alignment peaks seems to increase significantly. If, on the other hand, the second pulse is applied **at the time of half revival**, when the molecules are moving away from alignment, the torque impacted by the second pulse effectively **cancels the coordinated motion** of the rotating molecules, thwarting any future revivals. This is illustrated in figure 4b, where the second pulse was applied at $2.5T_{rev}$. These conclusions are also supported by other experiments in which molecular alignment was observed by weak field polarization technique[31] and by Coulomb explosion imaging[32].

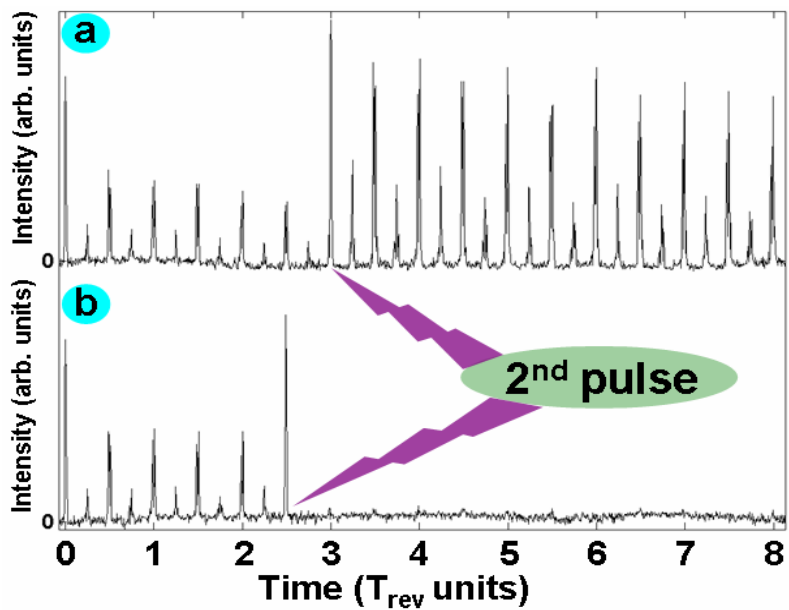

Figure 4: Alignment signal from $^{15}N_2$ gas (300 torr, room temperature), following excitation by two pulses. a) The two pulses are separated by a multiple of the exact revival time (3 $T_{rev}$). The torque from the second pulse adds coherently to that from the first one, resulting in observed enhanced alignment signal. b) The two pulses are separated by an odd multiple of half revival time (2.5 $T_{rev}$). The torque from the second pulse is opposite to the molecular angular velocity, resulting in effective stopping of the rotation.

___

## **Selective rotational excitation of Isotopologues**

Gaseous diffusion and centrifuge based isotope separation methods rely on mechanical effects caused by small mass differences. They are rather general, but relatively inefficient and require multiple separation stages. Laser isotope separation[33], on the other hand, provides high single-stage enrichment, but is molecule-specific, and since done in the frequency domain, it necessitates tunable narrow-band laser sources. A promising separation methodology[34,35,36,37] based on laser-induced vibrational wave packets in excited electronic states has been demonstrated, combining the advantages of the optical and mechanical methods, but it still relies on certain spectral selectivity. In what follows, we consider non-resonant isotope-selective rotational control in a mixture of different molecular isotopes of the same chemical element.

Following the application of an ultrashort laser pulse to a mixture of different molecules, a sequence of alignment and anti-alignment revivals is initiated. For similar species the periodicity is almost the same for all components of the mixture (no selectivity at this stage). However, with time different molecular "clocks" become desynchronized due to the difference in the moments of inertia (and rotational constants), and at well defined times, non-identical components (i.e. different isotopes) attain angular distributions which may be very different (i.e. cigar vs. disc).

We utilize the periodic behavior of the rotational wavepackets and the desynchronization discussed above for the specific addressing of molecular isotopologues too similar to be resolved on shorter time scales. As an example we use the homonuclear diatomic chlorine molecules, consisting of various mixtures of two natural isotopes $^{35}Cl$ and $^{37}Cl$, giving rise to three molecular species: $^{35}Cl-^{35}Cl, ^{35}Cl-^{37}Cl, ^{37}Cl-^{37}Cl$. Due to the (small) differences in their moments of inertia, the isotopic species will display very close revival periods. Consider two molecular isotopes with revival periods such that their (small) difference is given by $^{1}T_{rev} - ^{2}T_{rev} = \Delta\tau$. After one revival period the signals from the alignment peaks of the two isotopes will be separated by $\Delta\tau$, but after N cycles the difference will accumulate to $N\Delta\tau$, and the two species may be temporally resolved. In a set of experiments, we observed multiple revivals in Chlorine gas. The known rotational constants of these molecules correspond to a revival time of ~68 psec. At the first half and full revivals the signal from the different isotopes are too close to be resolved (Fig 5b), but at the second full revival (~140ps) (Fig.5c) the contributions of the three different isotopic species are separated by about 4 ps and are clearly and fully resolved. Note that based on the generic explanation leading to Figure 2, the signal from each isotope should have been a simple doubled peak. The deviation from this simple picture and the weaker 'extra' peaks arise from the centrifugal distortion of higher rotational states.

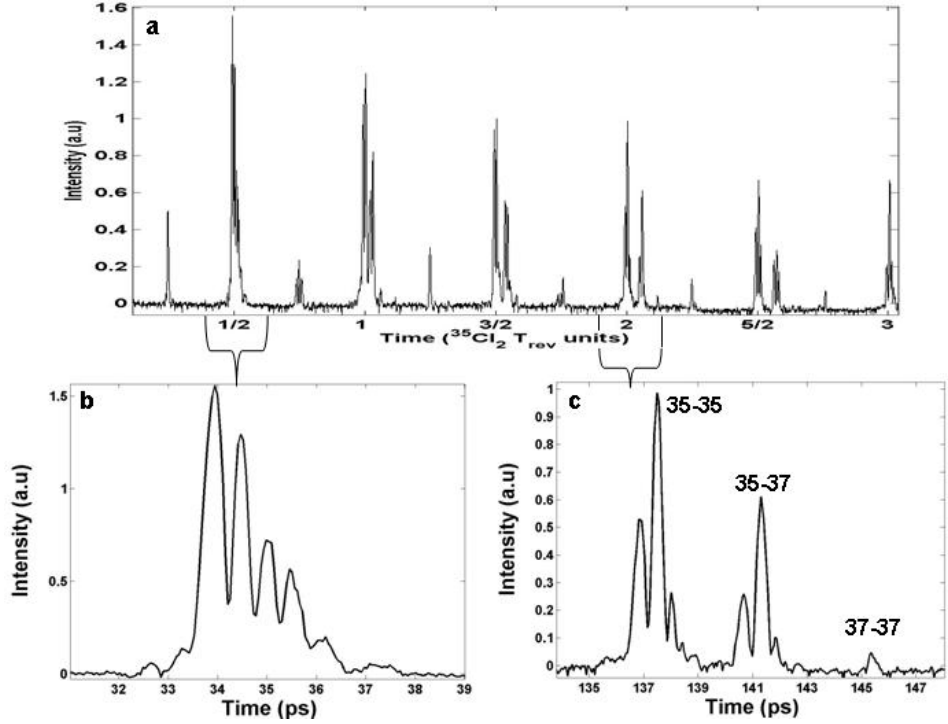

**Figure 5. a)** Long scan FWM signal observed from 225 torr $Cl_2$ gas. **(b)** Enlarged first half revival (~35 psec), the contributions of the different isotopic species temporally overlap. **(c)** Enlarged second full revival (~140 psec), the fully resolved contributions of the different isotopes are clearly seen.

In order to selectively manipulate a single species in a bimolecular mixture, we look for time windows where the contrast between the two components is maximal. Any two components of the mixture will have the maximum contrast of their angular distributions when one is aligned at exactly the same moment when the other goes through the state of anti-alignment. This contrast is found at times when one of the components completes an integer number of revival cycles while the second one performs "a half integer" number of its own cycles, i.e.

$$pT_{rev}^{(1)} = (q+\frac{1}{2})T_{rev}^{(2)}, \quad \text{or} \quad (p+\frac{1}{2})T_{rev}^{(1)} = qT_{rev}^{(2)} \tag{2}$$

where $p$ and $q$ are integers. As the ratio of the revival times $T_{rev}^{(1)}/T_{rev}^{(2)}$ for different isotopologues of the homonuclear diatomic molecule is a rational number, Eqs.(2) are linear Diophantine equations that may have solutions in positive integers $p$ and $q$. Thus, for nitrogen isotopologues $^{14}N_2$ and $^{15}N_2$, the revival time ratio is 14/15, and the suitable solutions are

$$(p,q) = (7,7), (21,22), (35,37),... \qquad (3)$$

Figure 6 depicts a measurement of the simultaneous alignment and anti-alignment of two isotopic components in a $^{14}N_2/^{15}N_2$ mixture as observed by the interference of their four-wave mixing (FWM) signals. Note that molecular alignment is reflected in the increase of the gas refractive index, while the anti-alignment causes its reduction compared to the isotropic case.

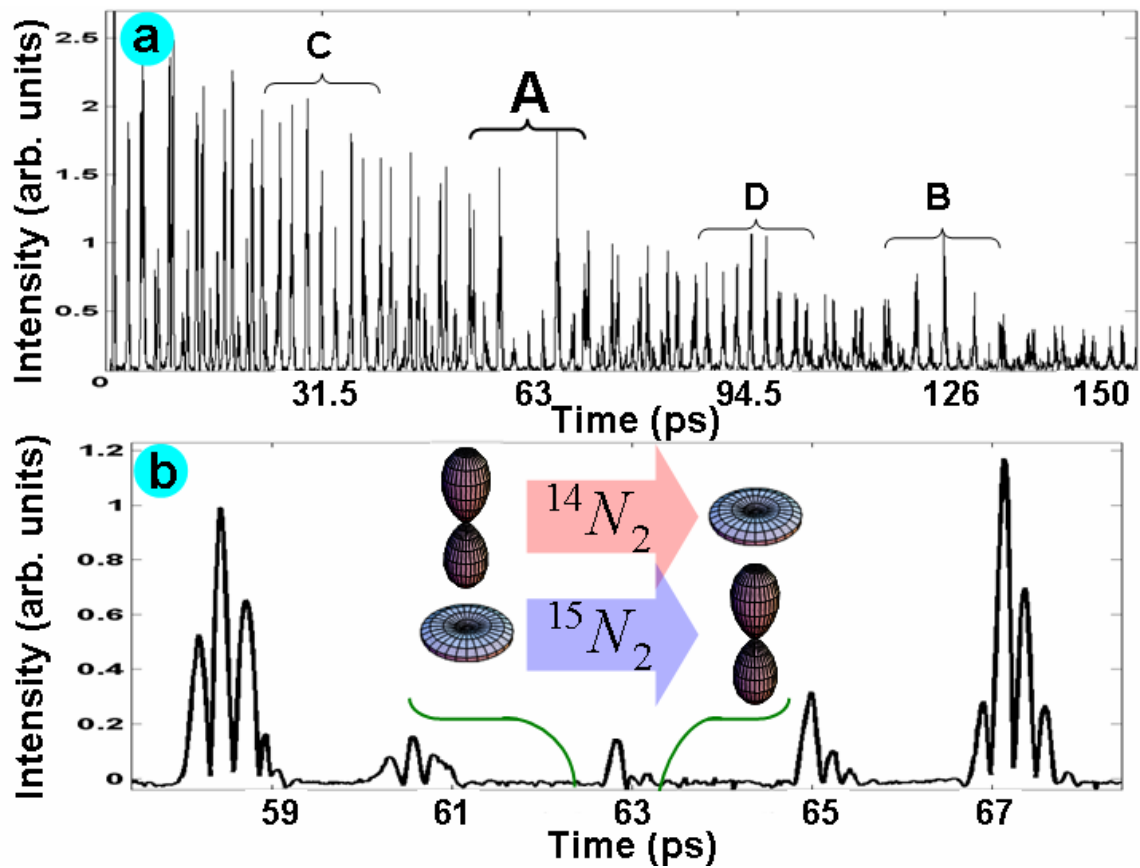

<u>Figure 6</u>: (color online) a) FWM signal from 1:1 $^{14}N_2 : ^{15}N_2$ mixture (500 torr, room temperature). Destructive interference of full and half revival signals is seen in region A at ~63 ps; Constructive interference of the two full revival signals is observed in region B at ~126 ps. Interferences of quarter and half revival signals are seen in regions C, D. b) Enlarged view of the destructive interference region A.

Figure 6a shows a full scan over many revivals of 1:1 mixture of $^{14}N_2, ^{15}N_2$. Compared to the single component picture (Figure 3) it shows a much more complicated envelope structure. The most profound features are a dip around ~ 63 ps (region A) and a peak at ~ 126 ps (region B). According to our previous analysis (see Eq.(3)), ~63 ps is the region where $^{15}N_2$ isotope completes 7 full revival periods while $^{14}N_2$ performs $7\frac{1}{2}$ of its own revival cycles. The reversed order of the alignment and anti-alignment events for these two isotopes causes a pronounced destructive interference in the combined FWM signal (see figure 6b). At this particular time, the sample experiences maximal angular separation of the isotopic components: when one of the isotopes reaches a cigar state, the other one exhibits a disc-like angular distribution, and vice versa. This provides a favorable configuration for further manipulation such as selective ionization (or dissociation) of the aligned component by an additional linearly polarized laser pulse. At ~126 ps, the $14^{th}$ full revival of $^{15}N_2$ and $15^{th}$ full revival of $^{14}N_2$ coincide, giving rise to constructive interference (region B). Other combinations of full and fractional revivals give rise to other interference phenomena (regions C, D), and these will be discussed in details in the next section, dealing with spin isomers.

We used this dramatic difference in angular distribution to achieve a two pulse isotope-selective control in the 1:1 mixture of $^{14}N_2$ and $^{15}N_2$. As detailed, at ~ 63 ps $^{14}N_2$ completes 7.5 revival cycles while $^{15}N_2$ completes 7 revival periods, one of the isotopes is rotating from the disc plane towards the cigar axis, while the other one goes in the opposite direction. A second pulse at this time would affect the two species very differently. As shown in figure 7, the second pulse at ~ 63 ps enhances the alignment of $^{15}N_2$ molecules, and almost completely stops the rotation of $^{14}N_2$ isotopes!

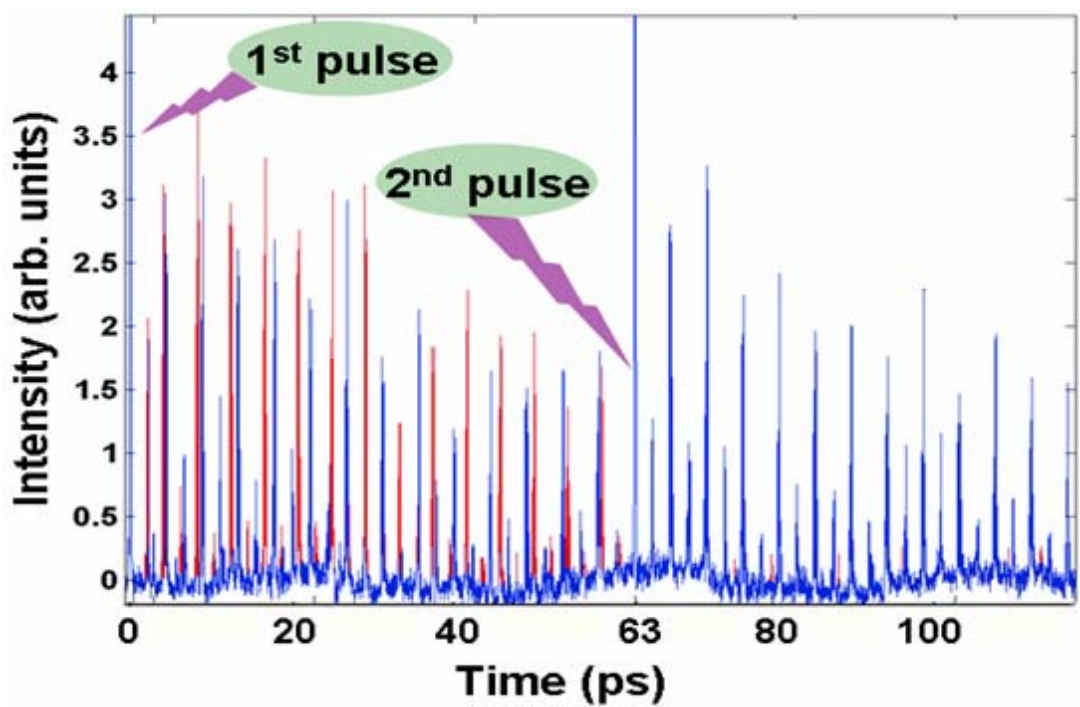

<u>Figure 7</u>: FWM signal from 1:1 mixture of $^{14}N_2$, $^{15}N_2$ (500 torr, room temperature) subject to two pulses ('kicks'), delayed by 63ps. The first kick excites both molecular isotopes, but the second kick, affects them in an opposite way. As a result, after the second kick, only the $^{15}N_2$ isotope experiences enhanced temporal alignment as reflected in the periodicity of the signal. The rotational excitation of the second isotope ($^{14}N_2$) is almost completely removed.

Up to this point, we have discussed the cases of mixtures of chemical species having **close** mechanical properties. We made use of the periodic evolution of their angular distributions and have shown that they can be distinguished spectroscopically with the aid of their rotational quantum revivals. We have demonstrated the selective rotational excitation using a double pulse scheme, of a single component and in bimolecular mixture, and briefly discussed the potential for physical separation of such mixtures by selective dissociation/ionization upon application of another laser pulse at specific times where the different components assume dramatically different angular distributions.

## The case of Spin Isomers

In the case of isotopologues we made use of the slight difference in the moment of inertia arising from the difference in masses of their atomic constituents. In the case of spin isomers, such a difference is not found. The chemical and physical properties of spin isomers are similar and in fact, the only difference between the spin isomers lies in their response to external magnetic field as in the case of Nuclear Magnetic Resonance. Therefore the ability to control their ratio in a mixture is not a simple task. A few practical applications of spin isomers are already known. For example, spin isomers can affect chemical reactions[38,39], can significantly enhance NMR signals[40,41] or can be used as spin labels in NMR based techniques.

Linear symmetric molecules whose atomic constituents posses a non zero nuclear spin exist as either Ortho or Para spin isomers, differing in the symmetry of their nuclear states. For a homonuclear diatomic molecule of spin ½ atoms, the total nuclear spin can be I=1 (Ortho molecule), or I=0 (Para molecule). This difference in symmetry implies statistical differences in the population of even and odd rotational states. Since spin changing collisions or radiational transfers are forbidden by symmetry, these species are rather stable, and once prepared can be used and stored for long times. At low temperatures, Ortho and Para molecules can be in principle separated based on the statistics they follow (Fermi-Dirac / Bose-Einstein) but in practice, this has been applicable only for the lightest of all molecules, Hydrogen.

Consider a spin ½ homonuclear diatomic molecule (e.g. $^{15}N_2$, $H_2$). Different spin quantum number combinations give rise to different symmetries of the rotational part of the wavefunction. For a spin ½ atom in a homonuclear diatomic, the spin symmetric molecular species populate the odd rotational states and the antisymmetric ones populate the even state as presented in Figure 8. In the case of a spin 1 atom, the correlations between the symmetry and the rotational state parity is reversed due the difference in the behavior of Fermions and Bosons. The difference in the parity of rotational states populated by the different spin isomers provides the key for their selective manipulation and in fact, the problem of selective excitation of spin isomers is reduced to selective excitation of even or odd rotational states.

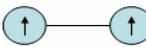

Figure 8: The four diatomic molecular spin isomer species, for a homonuclear diatomic molecule of spin ½ atoms. The spin function symmetry and corresponding rotational state parity for the spin ½ atom case are shown, from which one can see that the symmetric (odd J) species are three times more abundant.

In order to establish the connection between the *J* state parity and the angular distribution, we should remember that around full revival times all species behave similarly, evolving from antialigned "disk" state through the isotropic state to the aligned "cigar" state, and at half revival they follow the reverse order. As an illustration, the alignment factor ($<\cos^2\theta>$) during one full revival period for even and for odd states is depicted in figure 9.

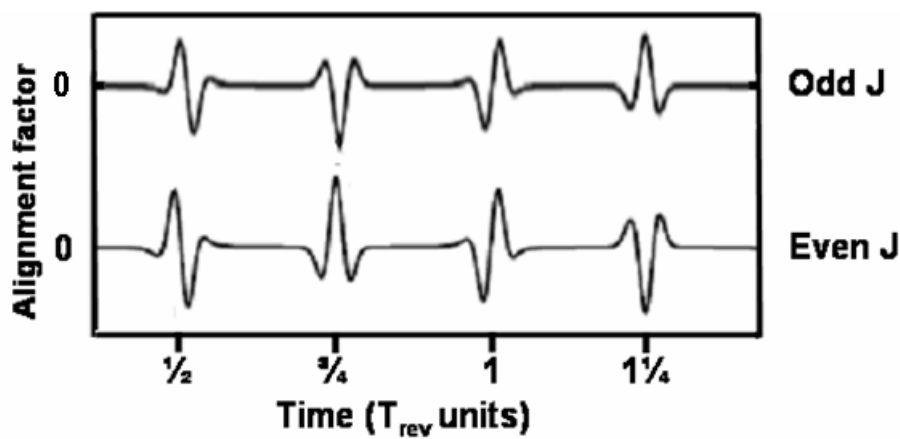

Figure 9: Schematic picture of the alignment factor ($<\cos^2\theta>$) along one revival period for the Even and Odd states separately.

As noted, around half and full revivals the odd and even states behave similarly, but around quarter revivals ( ¼ , ¾ ) they evolve in reversed orders such that around ¼ $T_{rev}$, the odd *J* states reach an antialigned ("disk") state while the even *J* states reach an aligned state ("cigar"). Around ¾ $T_{rev}$, they "switch their roles" such that the even *J*'s go through the disk state while the odd ones assume a cigar angular distribution. A second pulse applied around these times would affect the different parity rotational states (and corresponding spin isomers) in opposite ways. In a manner similar to the situation described in figure 4 above, the torque applied by the second pulse may either consent or oppose the already rotating molecules, depending on the exact timing and the sense of rotation.

We use the same experimental setup already described in the previous section. It is worth to note that since the excitation is a Raman type process ($\Delta J = \pm 2$), it maintains the *J* state parity distribution, meaning that a molecule initially occupying even (odd) states, will maintain it's state parity occupation after the interaction with the laser pulses. Since the experimental time domain data is analyzed in its Fourier space, we begin with a brief review of the main points of frequency domain CARS. A detailed theoretical analysis of the frequency domain is included in a forthcoming publication.

## Frequency domain analysis – test case

In FWM experiments the signal intensity is proportional to the square of the induced polarization, or in the language of transient gratings, the scattering intensity is proportional to the refractive index grating written by the two initial laser pulses. For this reason, both the aligned (high n) and the anti-aligned (low n) states will give rise to positive FWM signals. This is a marked difference from linear measurements where the actual sense of the change in refractive index may be observed, and positive as well as negative signals are observed Fourier analysis of this "positive definite" time delayed FWM signal results in an unusual frequency domain signature where binary sums and differences of the populated rotational states are observed.

As an example for this frequency domain analysis, let us examine the FWM signal from $^{15}N_2$ subject to a single excitation event.

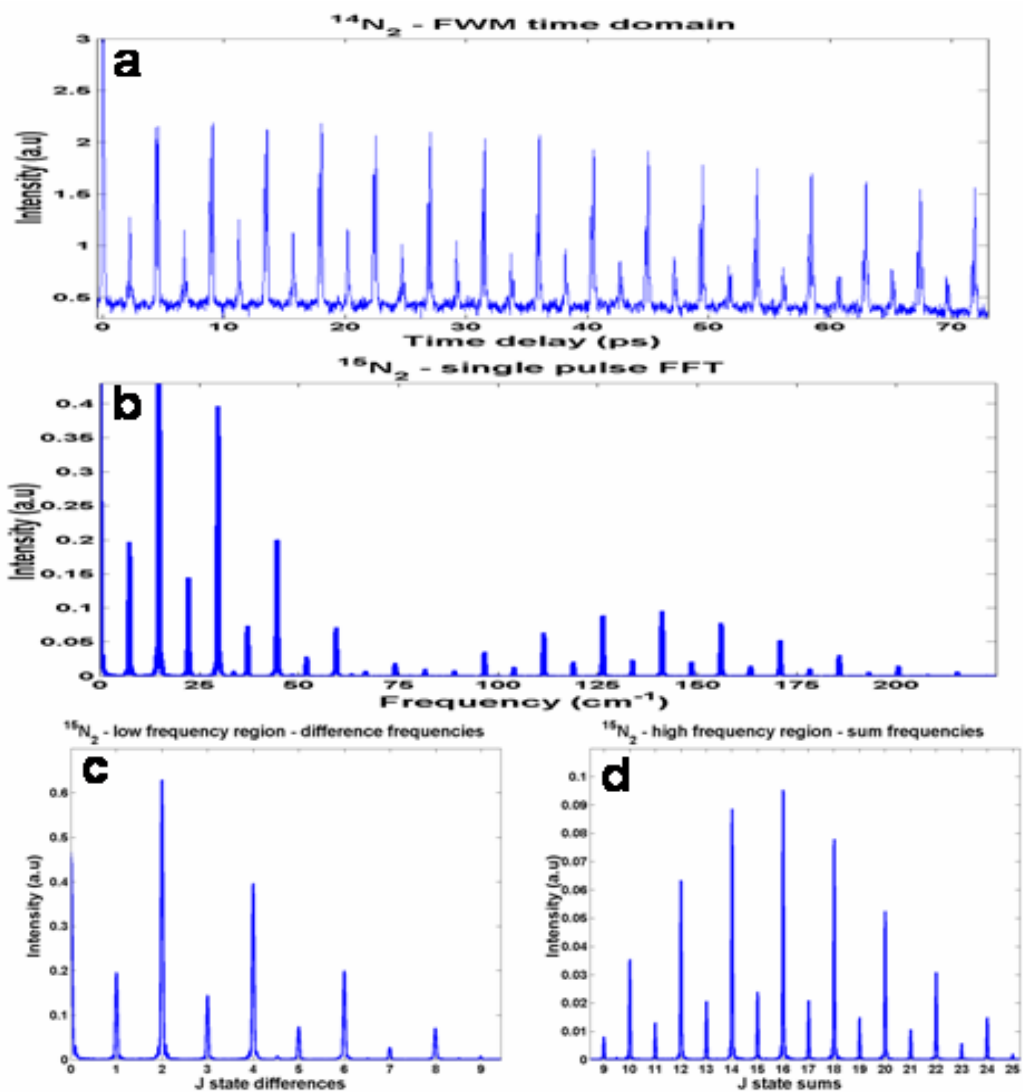

**Figure 10: FWM signals from $^{15}N_2$ applied by a single pulse. A) time domain, b) frequency domain, c) low frequency region – difference frequencies, d) high frequency region – sum frequencies.**

Figure 10a depicts the time delayed signal from $^{15}N_2$ subject to a single pulse and its full range Fourier Transform is depicted in Figure 10b. One can easily notice two regions, low and high frequencies corresponding to binary differences and sums of the populated $J$ states. Since the nuclear spin of $^{15}N$ is ½, the atoms should follow *Fermi-Dirac* statistics and the expected even/odd population ratio is: *$[J_{odd}/J_{even}] = (I+1)/I = 3:1$* .

In the low frequency regime (Figure 10c) we expect to find frequencies coming from binary **differences** of *J* states. If we denote such binary differences by *d*, one notes immediately that even *d* peaks are generated by *J* states of same parity whereas odd *d* peaks are generated by *J* states of different parity

Inspection of figure 10c reveals that the peaks at even *d* are higher then those at odd ones. Moreover, simplistically, one would have expected the *d=1* peak to be of the highest intensity (since it involves all neighboring *J* states), but instead, the *d=2* peak is stronger. Both these observations are explained by the fact that the even states are three times more populated then the odd states.

In the high frequency region (Figure 10d) we find frequencies arising from binary **sums** of the populated *J* states. If we denote such binary sums as *s*, one notes immediately that even *s* peaks are generated by *J* states of same parity whereas odd *s* peaks are generated by *J* states of different parity. As was the case of the low frequency region, here, too, the peaks at even *s* are higher, and for exactly the same reason.

Next we move forward to discuss double pulse schemes, where selective excitation is implemented.

## Selective excitations of Para/Ortho spin isomers

Our selective excitation scheme is based on the laser excitation of all components in the mixture, followed by wavepacket evolution and the application of a second pulse at a time when the angular distributions of the two species evolve in opposite directions. This happens around quarter and three quarters of the revival time for spin isomers.

We start with a comparison of single and double pulse excitations of $^{15}N_2$ in figure 11.

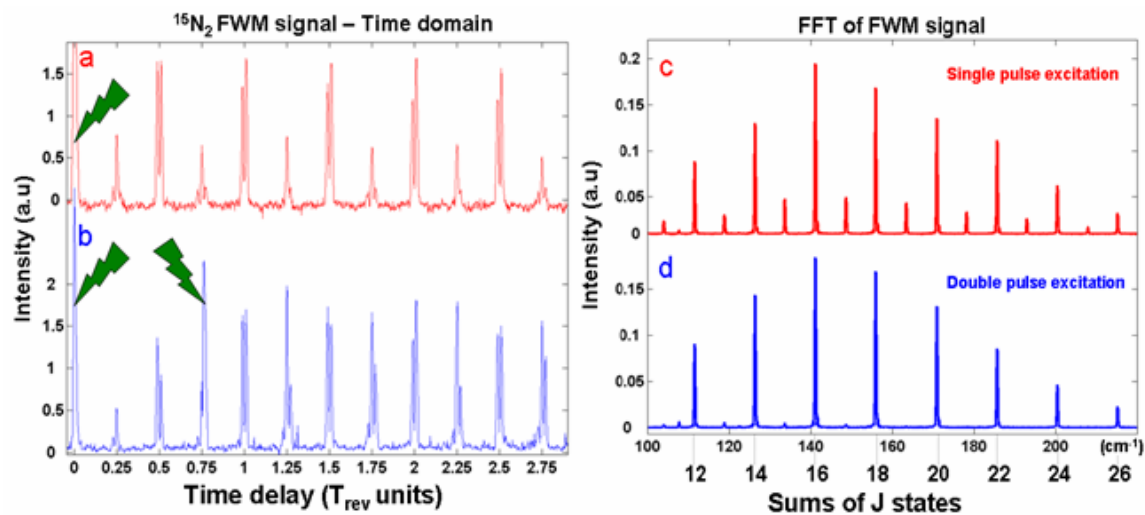

**Figure 11: a) Time domain FWM signal from $^{15}N_2$ excited by a single pulse. b) Time domain FWM signal from $^{15}N_2$ excited by 2 pulses delayed by $\sim\frac{3}{4}T_{rev}$. c) and d) Fourier transforms of a) and b) respectively, where the participating rotational states are clearly visible (see text).**

Figure 11a depicts the time domain FWM signal from $^{15}N_2$ following a single pulse at *t=0*. The horizontal axis is marked in units of revival times, so the large peaks correspond to full and half revivals, whereas the low intensity peaks at quarter revivals (¼ and ¾ $T_{rev}$) result from destructive interference of even and odd *J* states. In Figure 11b, a second pulse was applied around ¾ $T_{rev}$, at the right time to decrease the intensity of the even rotational states and enhance the rotation of the odd state. In 11b the interference is greatly reduced as is evidenced by the more or less constant intensity of the peaks. Note also that the intensity of the peaks observed *after* the second pulse is higher than the intensity of the peak *before* the second pulse. This is a consequence of the 3:1 odd/even population ratio in the case of $^{15}N_2$ (see figure 7 and the discussion that follows). Figures 11c,d depict the frequency region corresponding to binary sums of rotational states ("high frequency region"). The fact that only the even sums of *J* states are observed indicates that only states of single parity (even or odd), are coherently populated. The same effect is seen in the low frequency region, where only even J states' differences are observed (not shown here).

To find the accurate time when the two constituents are exactly out of phase with each other, we conducted two dimensional experiments where a full revival cycle was measured for a range of evolution times (given by the delay between the first two pairs of pulses). Figure 12a presents the time domain 2-dimensional plot where the horizontal axis is the probe delay, measuring the reviving rotational alignment peaks, and the vertical axis is the delay between the exciting pairs around the $T_{rev}= ¾$ time. Figure 12b is the line by line Fourier transform of 12a, showing the spectrum of excited rotational state (marked in actual rotational numbers) for the given evolution times on the vertical scale.

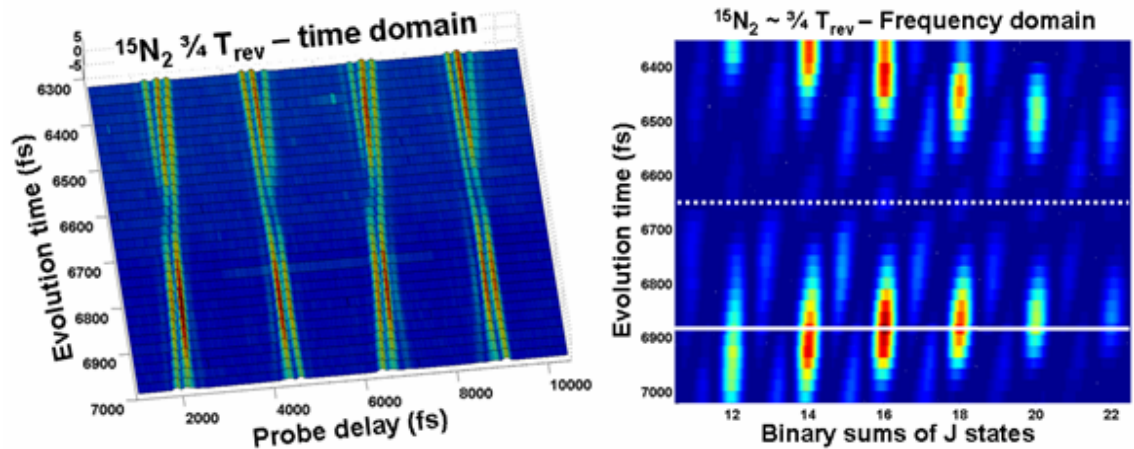

**Figure 12: 2D FWM scan for $^{15}N_2$. a) Time domain. The "evolution time" corresponds to time delay between the two excitation pulses. b) Frequency domain. As mentioned before, the even sums correspond to single J state parity excited (corresponding pure spin isomer excitation).**

The two horizontal lines in Fig 12b identify two specific time delays around the ¾ $T_{rev}$ (dashed, solid) where *J* states of single parity are excited (indicated by the even sums). The dashed line (intersecting the vertical axis at 6650 fs) corresponds to selective excitation of the even states (populated by only Para molecules), and the solid line (intersecting the vertical axis at 6880 fs) corresponds to selective excitation of odd *J* states (populated by only Ortho molecules). The intensity difference between the two time delays stems from the Ortho/Para population ratio (3:1 respectively). As discussed above, these unique time delays appear just before and after the exact time of alignment, when the angular distribution evolves from "disk" to "cigar" for enhancement (or in the reversed order for reduction) of the rotational energy. An equivalent behavior is observed for $^{15}N_2$ around $1¼\ T_{rev}$ (Figure 13a) with a reversed order of the molecular response.

These same effects are also observed for $^{14}N_2$. The nuclear spin of $^{14}N$ is 1, thus $^{14}N_2$ obeys Bose statistics and the resulting Para/Ortho (even/odd $J$ states) ratio is 2/1. The population ratio does not affect the qualitative behavior of the even and odd states but it is reflected in the details of the Fourier transform. This difference in nuclear spin also affects the correspondence between the $J$ states parity and the symmetry of spin isomers. Unlike the case of the fermionic $^{15}N_2$, in $^{14}N_2$ the symmetric spin wavefunctions occupy the even $J$ states and the antisymmetric ones occupy the odd ones. The bottom line of these differences is that the temporal behaviors of the two molecules are exactly complementary, as may be seen in figure 13. The figure depicts a comparison between the 2D Fourier transforms of these two species, where the interpulse delay in the double pulse excitation scheme was around ~$1¼\ T_{rev}$.

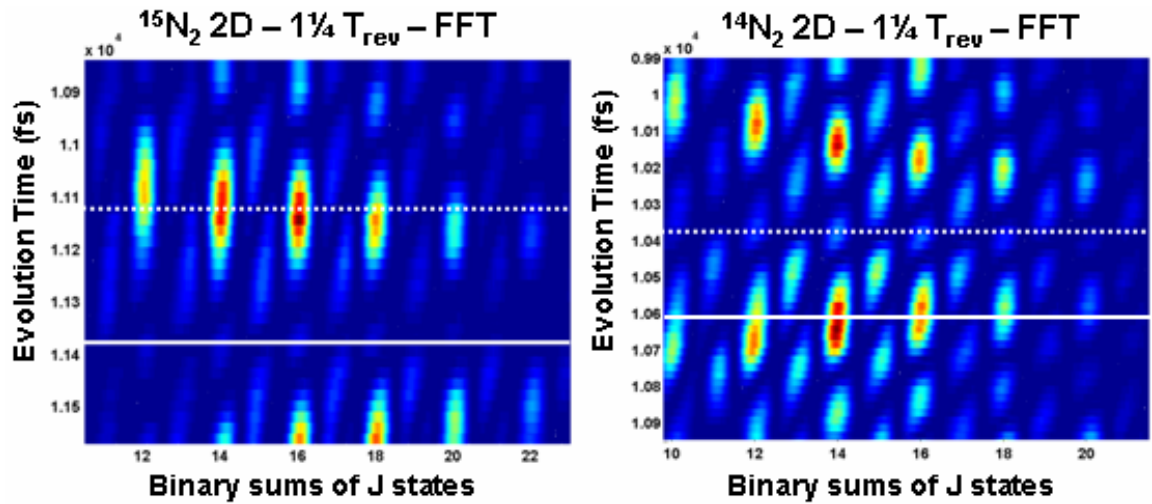

**Figure 13:** Comparison between 2D Fourier transforms of $^{14}N_2$ and $^{15}N_2$ applied by 2 pulses delayed by ~ $1¼\ T_{rev}$.

From the comparison between the 2D FFT of the $^{14}N_2$ (right) and $^{15}N_2$ (left), one can clearly see the difference in the ratio of peaks intensities at the dashed and solid lines. While in the case of $^{15}N_2$ the Ortho/Para (odd/even $J$'s) is 3:1, for $^{14}N_2$, the ratio of Para/Ortho (odd/even $J$'s) is 1:2. The behavior of the $J$ states remains the same but the relative population ratio and correlated spin isomers are changed. For a second pulse excitation which is applied around $3\backslash4\ T_{rev}$ (not shown), the intensities in each plot interchange with respect to the timing of the second pulse. Furthermore, if we compare the case of $^{15}N_2$, $1¼\ T_{rev}$ (figure 13 right picture) to the $^{15}N_2$, $¾\ T_{rev}$ (figure 12b) the intensities of the peaks are also reversed, as expected (see figure 9).

# Summary

In summary, multiple rotational quantum revivals are observed in molecular isotopic mixtures in response to an impulse excitation by a short pulse. By utilizing the repetitive nature of the alignment signal, we have shown that slight differences in the isotopes revival periods give rise to time-resolved discrimination between different isotopologues ($Cl_2$, $N_2$). Since the ratio between the isotopologues masses (and corresponding moments of inertia) is a rational number, one can easily calculate a time when the species to be separated reach a maximal difference between their angular distributions. At this time, one component evolves from disk to cigar (full revival time) distribution while the other evolves in the reversed direction (half revival time), setting the conditions for their selective excitation. A second laser pulse applied at that time will result in the alignment enhancement for one component and decrease for the other one. Around this specific time, dramatic difference exists in the angular distributions, and thus an opportunity arises for preferred ionization or dissociation of one or the other component. In the last part, we applied the double pulse scheme for selective excitation of spin isomers. In this case, there is no difference in the moment of inertia between the Para and Ortho isomers, therefore they revive at exactly the same times, and cannot be separated on a temporal basis. We have identified and demonstrated the difference in the behavior of Ortho and Para species around the quarter and three quarters revival times which is based on the different statistics they follow (Fermi-Dirac / Bose-Einstein) and leading to the maximal difference in their angular distributions around these times. Here, too, the second pulse affects the molecules in opposite ways; stopping the rotation of one component while enhancing it for the other.

The ability to selectively address a single species in a multi-component mixture, and change its physical properties (i.e. alignment, or rotation excitation level) is the main result of this work. Based on these observations, one may envisage ultrafast time-resolved analytical methods for isotope ratio determination, identification and discrimination of close chemical species, and trace analysis.

We acknowledge the support of the Israel Science Foundation, and the James Franck program at the Weizmann Institute.

# References

[1] M. A. Duguay and J. W. Hansen. Appl. Phys. Lett. **15**, 192 (1969).

[2] C. H. Lin, J. P. Heritage, and T. K. Gustafson. Appl. Phys. Lett. **19**, 397 (1971).

[3] C. H. Lin, J. P. Heritage, and T. K. Gustafson. Phys. Rev. Lett. **34**, 1299 (1975).

[4] For recent reviews on laser-induced alignment, see H. Stapelfeldt, T. Seideman. Rev. Mod. Phys. **75** (2003) 543; T. Seideman and E. Hamilton, Adv. At. Mol. Opt. Phys. **52**, 289, (2006).

[5] T. Seideman. Chem. Phys. **103**, 7887 (1995).

[6] J. Ortigoso, M. Rodriguez, M. Gupta, and B. Friedrich. J. Chem. Phys. **110**, 3870 (1999).

[7] F. Rosca-Pruna and M. J. J. Vrakking, Phys. Rev. Lett. **87**, 153902 (2001).

[8] I. Sh. Averbukh and R. Arvieu, Phys. Rev. Lett. **87**, 163601 (2001);
M. Leibscher, I. Sh. Averbukh, and H. Rabitz, Phys. Rev. Lett. **90**, 213001 (2003);
M. Leibscher, I. Sh. Averbukh, and H. Rabitz, Phys. Rev. A **69**, 013402 (2004).

[9] M. Renard, E. Hertz, S. Guérin, H. R. Jauslin, B. Lavorel, and O. Faucher, Phys. Rev. A **72**, 025401 (2005).

[10] C. Z. Bisgaard, M. D. Poulsen, E. Péronne, S. S. Viftrup, and H. Stapelfeldt, Phys. Rev. Lett. **92**, 173004 (2004).

[11] K. F. Lee, I. V. Litvinyuk, P. W. Dooley, M. Spanner, D. M. Villeneuve, and P. B. Corkum, J.Phys. B: At., Mol., Opt. Phys. **37**, L43 (2004).

[12] C. Z. Bisgaard, S. S. Viftrup, and H. Stapelfeldt, Phys. Rev. A **73**, 053410, (2006).

[13] D. Pinkham and R.R. Jones, Phys. Rev. A **72**, 023418 (2005).

[14] J. Karczmarek, J.Wright, P.Corkum, and M. Ivanov, Phys. Rev. Lett. **82**, 3420 (1999).

[15] I.V. Litvinyuk, Kevin F. Lee, P.W. Dooley, D.M. Rayner, D.M. Villeneuve, and P.B. Corkum. Phys. Rev. Lett. **90**, 233003 (2003).

[16] R.A.Bartels, T.C.Weinacht, N.Wagner, M. Baertschy, Chris H. Greene, M.M. Murnane, and H.C. Kapteyn, Phys. Rev. Lett **88**, 013903 (2002).

[17] V. Kalosha, M. Spanner, J. Herrmann, and M. Ivanov, Phys. Rev. Lett., **88**, 103901 (2002).

[18] R. Velotta, N. Hay, M. B. Mason, M. Castillejo, J. P. Marangos, Phys. Rev. Lett. **87**, 183901 (2001).

[19] M. Kaku, K. Masuda, and K. Miyazaki. Jpn. J. Appl. Phys. **43**, L591 (2004).

[20] J. Itatani, D. Zeidler, J. Levesque, M. Spanner, D.M. Villeneuve, P.B. Corkum, Phys. Rev. Lett. **94,** 123902 (2005).

[21] E. J.Brown, Qingguo Zhang and M. Dantus. J.Chem Phys **110**, 5772 (1999).

[22] M. Comstock, V. Senekerimyan, and M. Dantus. J. Phys. Chem. A **107**, 8271 (2003).

[23] The Encyclopedia of Separation Science, (ed.-in-chief I. Wilson), Academic Press (2000)

[24] For an early review, see B. Friedrich, D.P. Pulman, and D.B. Herschbach, J. Phys. Chem. 95, 8118 (1991)

[25] V. Renard et al. Phys. Rev. Lett. 90, 153601 (2003).

[26] J. H. Eberly, N. B. Narozhny, and J. J. Sanchez-Mondragon. Phys. Rev. Lett. **44**, 1323 (1980)

[27] Johathan Parker and C. R. Stroud, Jr.. Phys. Rev. Lett. **56**, 716 (1986).

[28] I.Sh. Averbukh and N.F. Perelman, Physics Letters, **139A**, 449 (1989)

[29] Y.Prior. App.Opt. 19, 1741 (1980).

[30] M. Leibscher, I. Sh. Averbukh, P.Rozmej, and R. Arvieu, Phys. Rev. A **69**, 032102 (2004).

[31] M. Renard, E. Hertz, S. Guérin, H. R. Jauslin, B. Lavorel, and O. Faucher, Phys. Rev. A. **72**, 025401 (2005).

[32] Kevin F. Lee, E. A. Shapiro, D. M. Villeneuve and P. B. Corkum, Phys. Rev. A. **73**, 033403 (2006).

[33] P.A. Bokhan, V.V. Bochanov, N.V. Fateev, M.M. Kalugin, M.A. Kazaryan, A.M. Prokhorov, D.E. Zakrevskii, "*Laser Isotope Separation in Atomic Vapor*", Wiley (2006)

[34] I.Sh. Averbukh, M.J.J. Vrakking, D.M. Villeneuve and A. Stolow, Phys.Rev.Lett. **77**, 3518 - 3521 (1996)

[35] M. Leibscher and I. Sh. Averbukh, Phys. Rev. **A63**,043407 (9), (2001)

[36] J.R.R Verlet, V.G. Stavros, H.H. Fielding, Phys. Rev. **A65**, 032504 (2002)

[37] A. Lindinger, C. Lupulescu, M. Plewicki, F. Vetter, A. Merli, S. M. Weber, and L. Wöste, Phys. Rev. Lett. **93** (2004), 033001.

[38] M. Quack: Mol. Phys. **34**, 477 (1977).

[39] D. Uy, M. Cordonnier, and T. Oka: Phys. Rev. Lett. **78**, 3844 (1997).

[40] C. R. Bowers and D. P. Weitekamp: Phys. Rev. Lett. **57**, 2645 (1986).

[41] J. Bargon, J. Kandels, and K. Woelk: Z. Phys. Chem. **180**, 65 (1993).